\documentclass[twocolumn,10pt]{asme2ej}

\usepackage{graphicx} 

\usepackage[]{appendix} 

\usepackage{cite}
\usepackage{url}

\usepackage{amsmath,amssymb,amsfonts}
\usepackage{algorithmic}
\usepackage{textcomp}

\usepackage[separate-uncertainty=true]{siunitx}

\usepackage{tikz}

\usepackage[ruled,vlined]{algorithm2e}

\usepackage[list=true,labelformat=simple]{subcaption}
\usepackage{xargs} 
\usepackage[colorinlistoftodos,prependcaption,textsize=tiny]{todonotes}
\newcommandx{\unsure}[2][1=]{\todo[linecolor=red,backgroundcolor=red!25,bordercolor=red,#1]{#2}}
\newcommandx{\change}[2][1=]{\todo[linecolor=blue,backgroundcolor=blue!25,bordercolor=blue,#1]{#2}}
\newcommandx{\info}[2][1=]{\todo[linecolor=green,backgroundcolor=green!25,bordercolor=green,#1]{#2}}
\newcommandx{\improvement}[2][1=]{\todo[linecolor=Plum,backgroundcolor=Plum!25,bordercolor=Plum,#1]{#2}}
\newcommandx{\thiswillnotshow}[2][1=]{\todo[disable,#1]{#2}}

\usepackage[firstpageonly=true]{draftwatermark}

\usepackage{hyperref}
\hypersetup{
    colorlinks=true,
    linkcolor=blue,
    filecolor=magenta,
    urlcolor=cyan,
    bookmarks=true,
}
\DraftwatermarkOptions{%
anchor=lt,
angle=0,
pos={50mm,50mm},
fontsize=0.012\paperwidth,
color={[gray]{0.2}},
text={
  \parbox{0.99\textwidth}{%
    \copyright\ 2023. This manuscript version is made available under the CC-BY 4.0 license\\ \href{https://creativecommons.org/licenses/by/4.0/}{https://creativecommons.org/licenses/by/4.0/}\\
    This is the accepted version of the manuscript. Published manuscript DOI: \href{https://doi.org/10.1115/1.4056105}{10.1115/1.4056105}%
  }},
}

\title{A Model-Free Kullback-Leibler Divergence Filter for Anomaly Detection in Noisy Data Series}

\author{Ruikun Zhou
    \affiliation{
	Department of Mechanical Engineering\\
	University of Ottawa\\
	Ottawa, Ontario, Canada\\
    Email: rzhou040@uottawa.ca
    }	
}

\author{Wail Gueaieb
    \affiliation{Member of the ASME\\
    School of Electrical Engineering\\ and Computer Science\\
	University of Ottawa\\
	Ottawa, Ontario, Canada\\
        Email: wgueaieb@uottawa.ca
    }
}

\author{Davide Spinello\thanks{Address all correspondence to this author.}
        \affiliation{Member of ASME\\
        Department of Mechanical Engineering\\
        University of Ottawa\\
        Ottawa, Ontario, Canada\\
        Email: dspinell@uottawa.ca
    }
}

\begin{document}

\maketitle    

\begin{abstract}
{\it We propose a Kullback-Leibler Divergence (KLD) filter to extract anomalies within data series generated by a broad class of proximity sensors, along with the anomaly locations and their relative sizes. The technique applies to devices commonly used in engineering practice, such as those mounted on mobile robots for non-destructive inspection of hazardous or other environments that may not be directly accessible to humans. The raw data generated by this class of sensors can be challenging to analyze due to the prevalence of noise over the signal content. The proposed filter is built to detect the difference of information content between data series collected by the sensor and baseline data series. It is applicable in a model-based or model-free context. The performance of the KLD filter is validated in an industrial-norm setup and benchmarked against a peer industrially-adopted algorithm.
}
\end{abstract}



\section{Introduction}\label{sec:introduction}

Anomaly detection generally refers to algorithmic procedures aimed at identifying relatively rare events in data sets. In the context of sensor-generated data series for inspection and monitoring, for instance, anomalies are typically defects to be detected. Anomalies can be operationally defined in terms of their relative information content with respect to a background data set characterizing anomaly-free noisy measurements~\cite{Khatkhate-Ray-Keller-Gupta-Chin-2006}. Several anomaly detection methods have been devised by leveraging tools from the field of information theory~\cite{wenke_lee_information-theoretic_2001,ghalyanSymbolicTimeSeries2020,liuGrowingStructureMultiple2009}. Entropy and relative entropy metrics are two main approaches in the family of information-theoretic anomaly detection techniques~\cite{chandola2009anomaly}. Entropy filters have been very popular in the fault detection of a wide range of systems~\cite{Dang-Chen-Wang-Ma-Ren-2020, Yu2020, Wang-Liu-Nunez-Dollevoet-2019}. Several entropy-based applications of machine fault-diagnostic systems are discussed in~\cite{Huo-Garcia-Zhang-Yan-Shu-2020}. An entropy filter has been proposed in~\cite{sheinker_magnetic_2008} to capture the anomaly in a magnetic field. An extension of the entropy filter presented in~\cite{spinello_entropy_2011} is proposed in~\cite{sheikhi_renyi_2015} to detect defects on the surface of gas pipelines 
The filter maps the raw data to a local information entropy space by assigning to each spatial location the entropy calculated from a neighborhood data set. A binary hypothesis test is then applied to partition the space between anomalous and background data series. 

The Kullback-Leibler divergence (KLD) is a measure of information entropy relative to a reference set, which is why it is also known as the relative entropy. This makes it a suitable candidate to partition and classify data. A number of studies have examined the effectiveness of the KLD in anomaly detection applications. Pietra et al.~\cite{della_pietra_inducing_1997} applied it to natural language processing. The KLD is also used to detect anomalies in network security applications based on data traffic analysis~\cite{gu_detecting_2005}. In~\cite{afgani_information_2008}, it is used to capture signal features of cognitive radio. In addition, the KLD has been applied for anomaly detection in periodic signals~\cite{afgani_anomaly_2008,afgani_hardware_2009,afgani_exploitation_2011}. These studies can be classified as model-free since they stochastically represent the data through discrete random variables without assumptions on the underlying probability distributions. In applications involving signals of known nature, the data is often fitted to a specific statistical model. Such an approach may lead to computationally advantageous filters by leveraging the statistical properties of the adopted models. For example, a Gaussian approximation has been adopted in~\cite{sheikhi_renyi_2015} to construct an entropy filter for anomaly detection, and in~\cite{zeng_detecting_2014,xie_fault_2015,  youssef_optimal_2016, chen_improved_2018, jabari_multispectral_2019} for KLD-based algorithms. Based on this analytical treatment, a more recent work by~\cite{bouhlel_kullbackleibler_2019} derived the closed-form of KLD for multivariate generalized Gaussian distributions. Within the Gaussian approximation, a KLD algorithm for monitoring statistics is presented in~\cite{zeng_detecting_2014}, while in~\cite{harmouche_incipient_2016} the KLD measure is used to find the incipient fault in multivariate processes by taking advantage of its high sensitivity to small changes in signals, where the fault magnitude is also accurately estimated by relative KLD values. The KLD is also applied in~\cite{youssef_optimal_2016} for crack detection in eddy current testing, while~\cite{hamadouche_modified_2017} addresses the fault detection in large scale systems based on a non-parametric approximation with a modified KLD.  A recent work by~\cite{chen_improved_2018} extended this technique to the analysis of online data with off-line references under a multivariate statistical analysis framework. The application of the KLD in mechanical-type fault diagnosis is explored in~\cite{mezni_bearings_2018} for bearing balls. Moreover, some scholars have studied the advantages of KLD compared to conventional multivariate statistical approaches for anomaly detection. For instance, a comparison between the KLD measure and principal component analysis is made in~\cite{xie_fault_2015}, where it is shown that the sensitivity of the KLD measure is superior to the one from the principal component analysis.

The contribution of this paper is the formulation of a KLD filter for anomaly detection in noisy data series with two main advantages. The first is the filter's flexibility to be applied in a model-based or a model-free context, in the sense that sampled data may or may not be fitted in a priori known probability distribution model. This is important because fitting the data into a pre-defined model requires an offline sampling of the data beforehand, and even so finding a closed-form probability distribution that best fits it is not a trivial process. The second advantage of the proposed filter is its threshold-free nature, which means that it does not depend on a pre-determined threshold to discriminate defected from healthy data samples. This is convenient because deciding on a suitable pre-defined threshold can be tedious and time consuming. 

Unlike some similar techniques~\cite{Abid-Khan-Iqbal-Silva-2021}, the KLD-based algorithm presented here is not predicated on assumptions about the data distribution, and therefore it can be implemented in a model-free framework. However, it can be adapted to include distribution models whenever they are reliably available for a specific system or sensor. The possibility of applying the KLD in a model-free approach on fault detection has been discussed in~\cite{youssef_optimal_2016}, with different simulated scenarios showing its effectiveness in detecting the location of cracks from eddy current sensor data. Previously, generalized likelihood ratio-based and KLD-based tests have been shown to be effective alternatives to more traditional statistical tests for incipient fault detection~\cite{tanakaFaultDetectionLinear1990a, harmouche_incipient_2014}. Since the KLD intrinsically measures the discrepancy in information content between two data sets, it is very robust in anomaly detection as it does not require a priori knowledge about the nature of the anomaly. Furthermore, it does not require accurate statistical characterization of the data~\cite{1621202,anderson_kullback-leibler_2011}, or prior offline learning~\cite{Abid-Khan-Iqbal-Silva-2021, Atienza-Bielza-Diaz-Larranaga-2021}. More importantly, contrary to other approaches reported in the literature, such as in~\cite{youssef_optimal_2016}, the proposed algorithm is not only able to find the locations of the anomalies, it can also identify their relative sizes as well.

\section{KLD filter} \label{kld filter}

\subsection{Preliminaries on Information Theory} \label{KL divergence}

In this work, we focus on data series produced by a family of digital sensors where statistical constructs can be defined as discrete random variables. This allows to adopt a robust model-free approach as it can accommodate for signals that substantially depart from the usually assumed Gaussian distribution, or more generally for signals that cannot accurately be represented by uni-variate or bi-variate distributions, being multi-variate in nature. Therefore, the model-free approach allows a broader scope and applicability of the proposed method. Let $X$ be a discrete random variable mapping to a finite dimensional event space $\{x_1,x_2,\ldots,x_N\}$ of cardinality $N$. Let $P(x_i)$ be a probability mass function over the set $X$ with $\sum_{i=1}^N P(x_i) = 1$. The information entropy $H$, or Shannon entropy, is the expected value of the information content $-\log P(X)$ of the random variable $X$~\cite{shannon1948mathematical},
\begin{align}\label{eq_entropy}
  H(X) &:= - \sum_{i=1}^N P(x_i) \log P(x_i)
\end{align}
The relative entropy $D_{\rm KL}(P || Q)$, or KLD, is the expected value of the logarithmic difference (or information divergence) between two distributions $P$ and $Q$ defined over the same random variable $X$~\cite{kullback1951information}, with the expectation taken relative to~$P$, 
\begin{align}\label{eq_KL_D}
    D_{\rm KL}(P || Q) &:= \sum_{i=1}^N P(x_i) \log \frac{P(x_i)}{Q(x_i)}
\end{align}
It is important to notice that the KLD is not symmetric since $D_{\rm KL}(P || Q) \neq D_{\rm KL}(Q || P) $, and therefore it has to be intended as a measure of discrepancy between the two distributions rather than a metric. Also, the KLD is a non-negative real number for specific distributions and it evaluates to zero only if $P=Q$~\cite{kullback_information_1959}. In the following, we will adopt the convention that $0\log\frac{0}{Q} = 0$ when $P(x_i) = 0$,  $P\log\frac{P}{0} = \infty$ when $Q(x_i) = 0$, and $0\log\frac{0}{0} = 0$~\cite{cover_elements_2006}.

\subsection{Anomaly Detection Algorithm} \label{kld_modelfree}

\begin{figure} [!htb]
	\centering
	\includegraphics[scale=0.8]{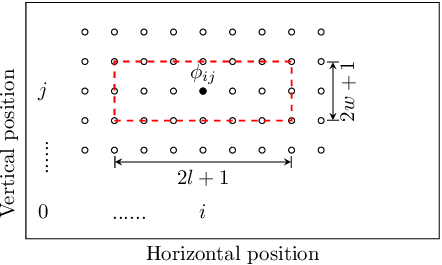} 
	\caption{Schematic of data points on a 2D surface, where the dashed rectangle represents a window centered at datum $\phi_{ij}$ .}
	\label{fig_window}
\end{figure}

The KLD filter in this work maps each raw data point to a local KLD value. To illustrate the core idea, consider a 2D array of sensor measurements representing uniformly spatially scattered data points (or scalar fields), as in the case of projecting a Lidar onto a wall or sliding a proximity sensor across the inner surface of a pipe, for example. This is schematically illustrated in Fig.~\ref{fig_window}, where a local portion of the sensed surface is represented in Cartesian coordinates. This would also apply to cylindrical surfaces (e.g.,~a pipe) by converting the cylindrical coordinates of the data points to Cartesian coordinates. Each data point on the surface, representing a measurement of the scalar field $\phi$, is identified by two integer indices, $i = 1,\ldots,N_\ell$  and $j=1,\ldots,N_w$, respectively, spanning the horizontal and the vertical directions. Therefore, the raw measurement of the field $\phi$ is denoted by $\phi_{ij}$. The data set comprised of all measurement data is denoted by~$\mathbb{S}$.

The KLD as expressed in~\eqref{eq_KL_D} can be intended as a weighted likelihood test, where the hypothesis of a data point $\phi$ being an anomaly is tested against the null hypothesis of $\phi$ being an expected measurement. In this sense, the distribution $Q$ of the baseline expected signal can be regarded as the prior, whereas the distribution $P$ of the sensor measurements can be regarded as the posterior. Then, the anomalies can be identified by the discrepancy between these two distributions. Meanwhile, a well known problem in the calculation of the KLD for discrete data series may arise in~\eqref{eq_KL_D} when the denominator $Q(X)=0$ while the numerator $P(X) \neq 0$. This is an occurrence to be expected if $Q$ is built from a noisy data that is substantially distinguished from the anomalies. To overcome this problem, we build the histogram and probability mass function $Q$ from the entire data set. If the anomalies are relatively rare, the distribution of $Q$ is still dominated by the noise around the expected measurement, therefore providing a suitable approximation that does not suffer from empty bins. A measurement point corresponding to an anomaly evaluates to a small probability $Q(\phi) \ne 0$, therefore overcoming this issue. On the other hand, the local nature of the proposed algorithm biases the distribution $P$ towards anomalies if any is present within the data subset used to build it.

\begin{figure*}[h!t]
	\centering
	\subcaptionbox{Raw sensory data series dominated by background noise in the presence of anomalies.\label{fig_0degree}}{\centering
		\includegraphics[width=0.48\textwidth, angle = 0]{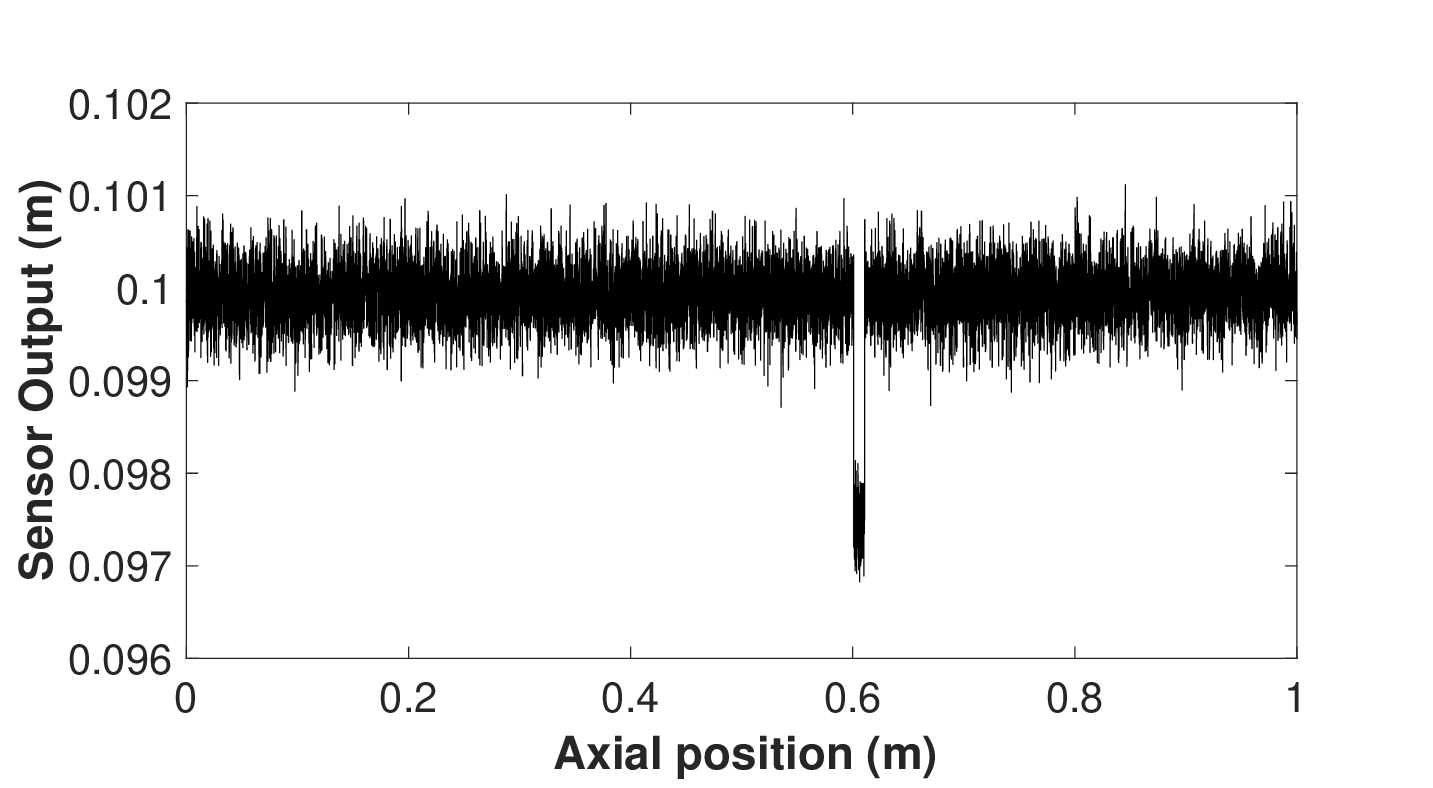}}
   \hfil
	\subcaptionbox{Raw sensory data series with noise only.\label{fig_noisy}}{\centering
		\includegraphics[width=0.48\textwidth, angle = 0]{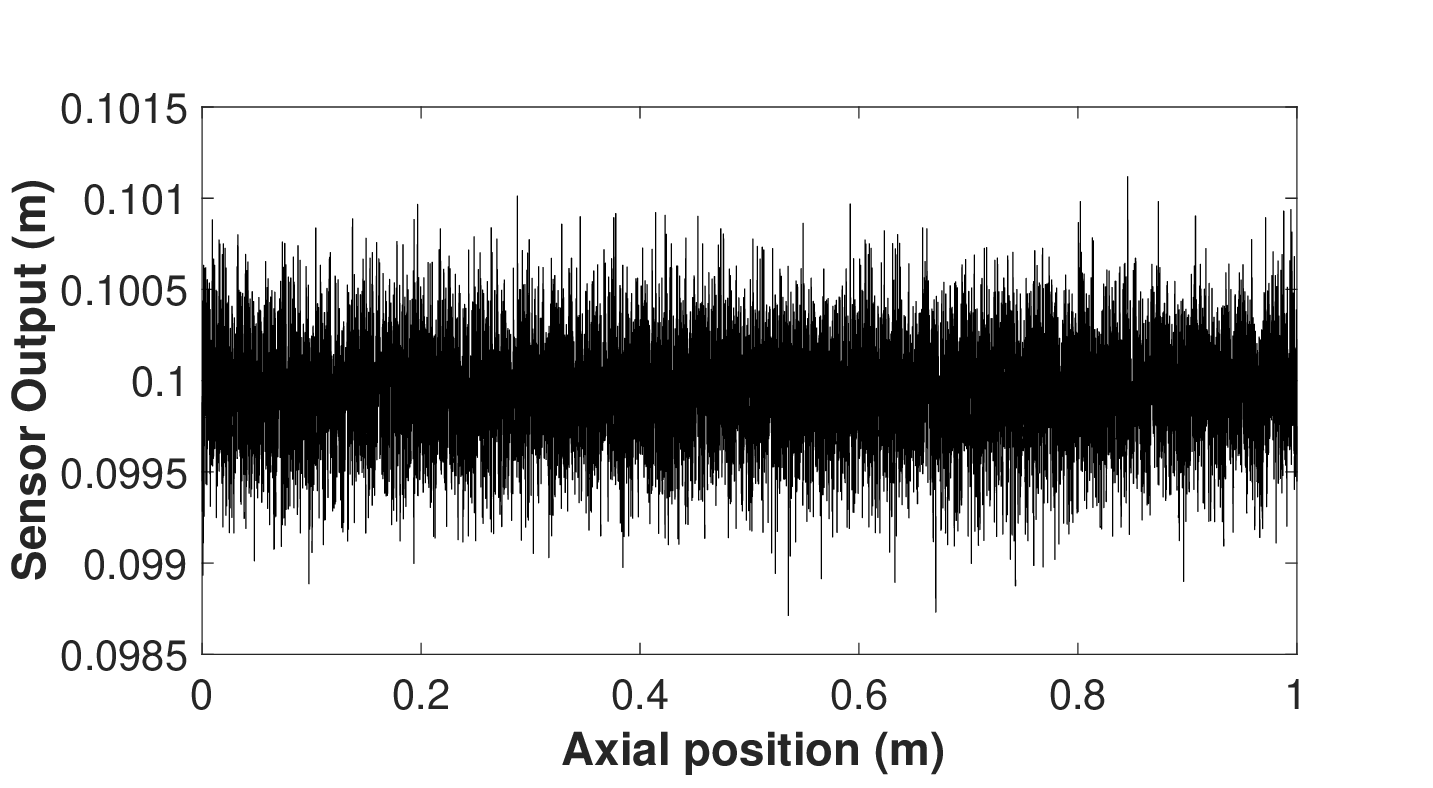}}
              \caption{Raw data series with and without anomalies.}
	\label{fig:rawData}
\end{figure*}

To illustrate the steps involved in building the KLD filter, we consider a one-dimensional data series of sensory samples collected by displacing a proximity sensor over a line segment. This type of sensor is only considered here for illustrative purpose. The algorithm is applicable to other types of sensors too. We will assume that the data series includes a few anomalies, which represent rare events in the data dominated by the background noise (see Fig.~\ref{fig_0degree}). The purely noisy data can be found in Fig.~\ref{fig_noisy}. By comparing these two figures, we can conclude that the anomalies can be definitely regraded as rare events. And it is possible to compute the KLD values for each data point by using the noisy data as the prior, but some pre-processing steps will be needed to avoid the zero denominator case.

\begin{figure*}[h!t]
	\centering
	\subcaptionbox{Histogram of the data in the data series in Fig.~\ref{fig_0degree} used to build the baseline probability distribution $Q$.\label{fig_his_q}}{\centering
		\includegraphics[scale=0.35, angle = 0]{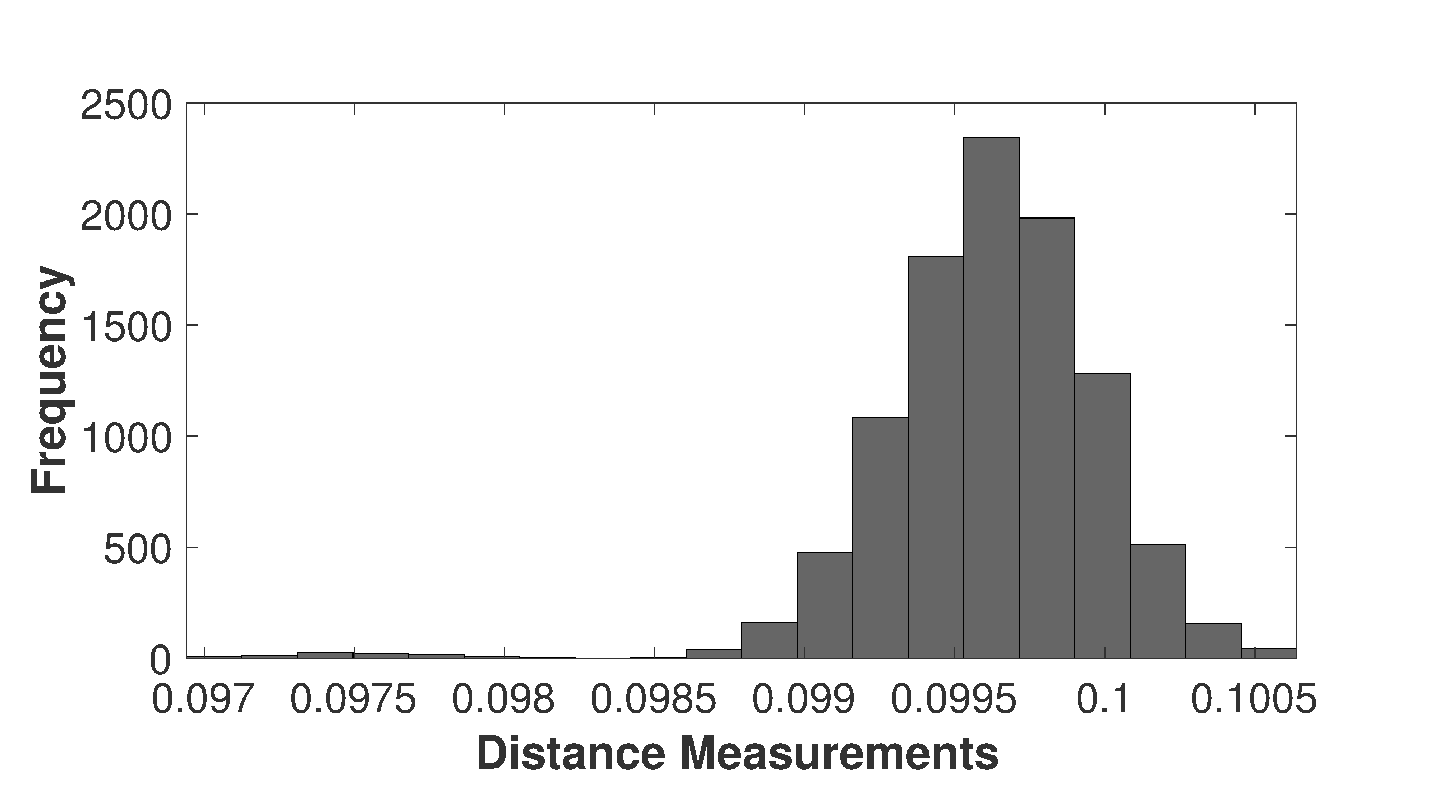}}
   \hfil
	\subcaptionbox{Probability mass function.\label{fig_pmf_q}}{\centering
		\includegraphics[scale=0.35, angle = 0]{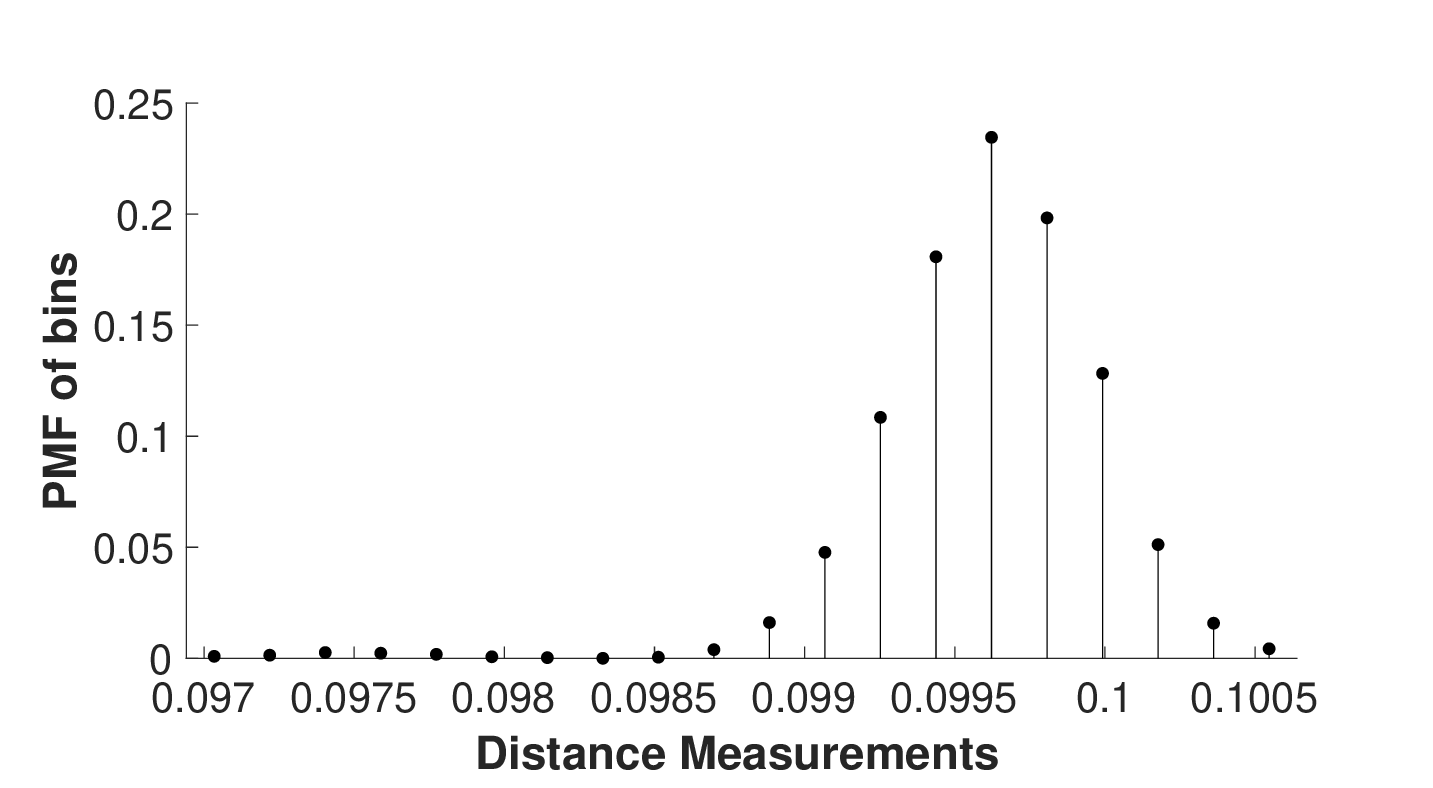}}
              \caption{Histogram and probability mass function of the baseline distribution $Q$ in the KLD calculation.}
\end{figure*}

First, the mass function $Q$ is obtained by building a histogram of all the data set $\mathbb{S}$ with $k$ bins. Denoting the minimum and the maximum values by $a = \min (\mathbb{S})$ and $b = \max (\mathbb{S})$, respectively, and restricting the bins to have an equal size $h$, we get $h=(b-a)/k$. Let $B_n = [a+(n-1)h, a+nh)$ denote bin~$n \in \{1, 2,..., k\}$, bounded by $a+(n-1)h$ and $a+nh$. Fig.~\ref{fig_his_q} shows an example of the histogram with 20 bins. Based on the histogram, the corresponding probability mass function (PMF), denoted by $Q(\phi)$, is obtained by normalizing the frequencies in every bin with respect to the cardinality of the data set, as shown in Fig.~\ref{fig_pmf_q}. We observe that the discrete values taken by $Q$ represent the probability that a measurement $\phi$ belongs to the corresponding bin of the histogram used to generate $Q$. Therefore, the mass function can also be used to evaluate probabilities of the original data set. The PMF shows the majority of the events happening around the expected measurement of the noisy sensor signal, with a tail generated by the anomaly, as expected considering that it is a rare event. We do have zeros in $Q(\phi)$, in this case, but at that point the corresponding probabilities in $P(\phi)$ are zero as well, which returns zero terms in the summation for the calculation of the KLD, as per the convention stated earlier, i.e.,~$0\log\frac{0}{0} = 0$.

Next, we need to map each raw data point $\phi_{ij}$ into a local KLD value. To do so, we consider two integer parameters, $l$ and $w$, to define a window of size $(2 l + 1) \times (2 w + 1)$ centered around the location $(i,j)$, as illustrated in Fig.~\ref{fig_window}. The window delimits a rectangular region enclosing a subset $\Phi(i,j) \subsetneq \mathbb{S}$ of cardinality $(2l+1)(2w+1)$. From each data subset $\Phi(i,j)$ associated to the raw data point $\phi_{ij}$, a probability mass function is built by considering the samples in $\Phi(i,j)$ as a random variable ${\Phi_{ij}}$ whose realizations are the data points in the window centered at $\phi_{ij}$. This allows to build a probability mass function for each data point, so that the map into the KLD has a local nature, which allows to detect spatial variations of the data~\cite{sheikhi_renyi_2015}. 

One of the advantages of this algorithm is that it requires no prior information about the anomalies. Since the prior distribution $Q$ is built from the entire data set $\mathbb{S}$, which includes each subset $\Phi(i,j)$, for every sample $\phi \in \Phi(i,j)$ the probability $Q(\phi)\ne 0$, and the calculation of the KLD is well posed. The baseline distribution $Q$ could be built from a different data set, for example from typical noise data characterizing the same class of sensors as the one used. This would have the advantage of not relying on data from the same sensor for the baseline distribution, therefore allowing online implementations of the filter. However, it could generate scenarios in which $Q(\phi)=0$ while $P(\phi)\ne0$ for the same measurement $\phi$. An approach adopted when there are few points with zero probability $Q$ is to simply ignore the empty bins and implicitly set the corresponding terms to zero, namely, ignore these terms in the summation without significantly affecting the result if the number of empty bins for $Q$ is small relative to the total number of bins~\cite{murphy_machine_2012}. Another possibility is to fictitiously add the same constant small frequency to all bins, so that no empty bins exist~\cite{afgani_anomaly_2008}, or by applying a boot-strapping method (Monte Carlo approach) for filling up the missing data if there are any empty bins caused by measurement missing. If the number of empty bins for $Q$ is significant, a more rigorous approach is to adopt Jenson-Shannon divergence~\cite{callegari2017information}. We stress one more time that as long as the anomalies are rare, $Q$ is expected to be almost fully characterized by noise around the expected measurement, and therefore we can avoid the indeterminacy caused by $Q(\phi) = 0$, while the local KLD would measure the local discrepancy between a data point and the overall data set which closely resembles noise. The analysis of the error from the approximation of using the entire data set as baseline is done in the next section.

\begin{figure*}[ht!]
	\centering
	\subcaptionbox{Histogram of $\Phi(i,j)$ when $\phi_{ij}$ is a data point from the series in Fig.~\ref{fig_0degree} located on the anomaly, with $k=20$ bins.\label{fig_his_p}}{\centering
		\includegraphics[scale=0.35, angle = 0]{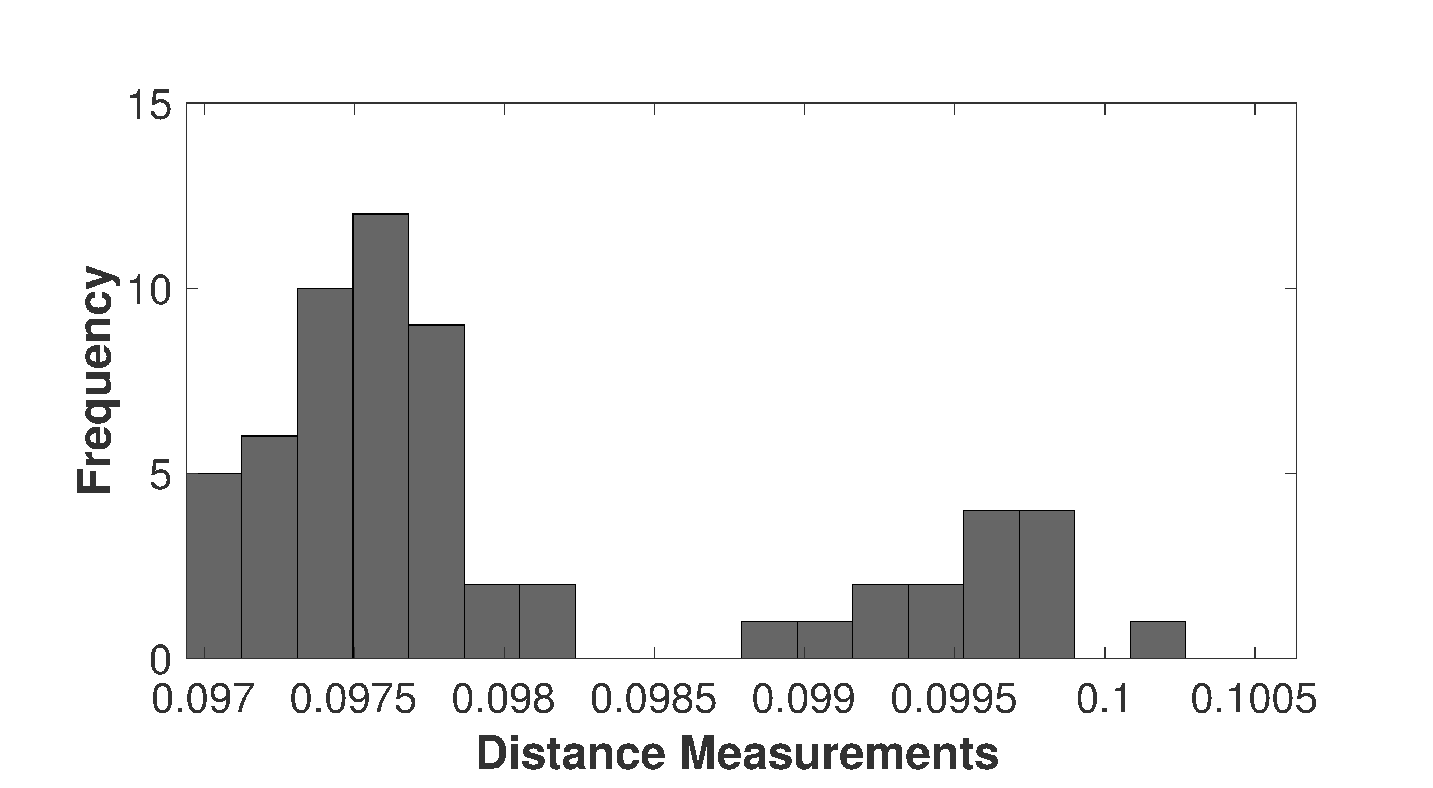}}
   \hfil
	\subcaptionbox{Probability mass function.\label{fig_pmf_p}}{\centering
		\includegraphics[scale=0.35, angle = 0]{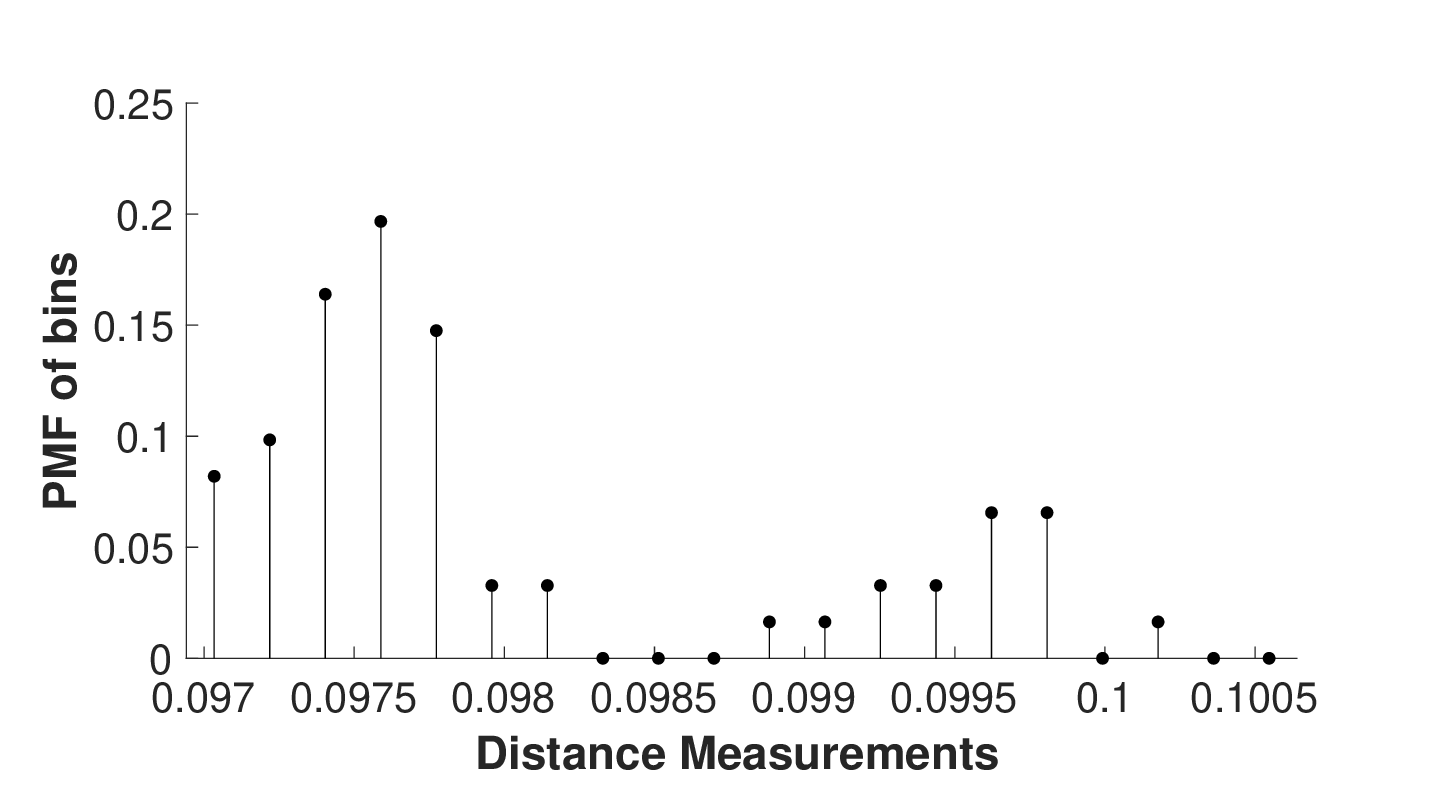}}
              \caption{Probability mass function with 20 bins.}
\end{figure*}

It is interesting to notice that the histogram of $P_{ij}$ significantly differs from that of $Q$ (by comparing Figs.~\ref{fig_his_q} and~\ref{fig_his_p}). This is expected since most events in $P_{ij}$ fall into the first few bins due to the fact that the anomaly corresponds to a smaller measurement. As a result, it appears with a high frequency since the window used to build the histogram captures a significant amount of data related to anomaly measurements in the first few intervals. On the other hand, the histogram of $Q$ in Fig.~\ref{fig_his_q} is dominated by the higher measurement values that are representative of the sensor's output corresponding to non-anomalous data.

The last step in building the filter is to measure the local discrepancy between the prior distribution $Q$ and the posterior $P_{ij}$, $\forall i,j$, which results into mapping each raw data point $\phi_{ij}$ into a local KLD
{\allowdisplaybreaks \begin{align}\label{eq:KL_div_loc}
D_{\rm KL}(P_{ij} ||Q) &= \sum_{\phi \in \Phi_{ij}} P_{ij}(\phi) \log \frac{P_{ij}(\phi)}{Q(\phi)} \nonumber
\\
&= \sum_{n=1}^{K} P_{ij}(\phi \in B_n) \log \frac{P_{ij}(\phi \in B_n)}{Q(\phi)}
\end{align}}%
where $B_n$ is the $n$-th bin in the posterior $P_{ij}$ and $K$ being the total number of bins. Expression~\eqref{eq:KL_div_loc} greatly simplifies the calculation of the KLD, since the summation extends over the bins rather than the entire data samples in $\Phi(i,j)$. However, we need to clarify the meaning of the expression $Q(\phi)$, since the bins used to build the histogram that generates $Q$ can in general be different than the ones used for the histogram that generates $P_{ij}$. Let $\delta_{ij}$ be the uniform bin size of the histogram of $\Phi(i,j)$, and $h$ be the bin size for the histogram of $\mathbb{S}$, as defined earlier. We define the discrete probability density functions $p_{ij}$ and $q$ by normalizing each value of the corresponding mass functions by the bin sizes, so that 
\begin{align}
P_{ij}(\phi \in B_n) & = p_{ij}(\phi_n)\delta_{ij} & Q(\phi) & = q(\phi_n)h
\label{eq:pq_bins}
\end{align}
where $\phi_n$ is any element in bin $B_n$, which can be chosen arbitrarily within the bin, since by construction, all the data points in a bin have the same probability. Note that, in general, the bin size for the histogram generating $Q$ is different. That is $k \ne K$. However, $q(\phi_n)$ accounts for this since the mass $Q$ is normalized by the bin size. Substituting into~\eqref{eq:KL_div_loc}, yields
\begin{align}
D_{\rm KL}(P_{ij} ||Q) = \sum_{n=1}^{K} p_{ij}(\phi_n) \delta_{ij} \log \frac{p_{ij}(\phi_n) \delta_{ij}}{q(\phi_n)h}
\label{eq:KL_div_loc2}
\end{align}
Expression~\eqref{eq:KL_div_loc2} allows to use a different number of bins and different bin sizes for data sets that may have different structures. As the summation is extended only to the bins for the prior $P_{ij}$, and as $\Phi(i,j) \in \mathbb{S}$, the relationship can be rewritten as
\begin{align}
D_{\rm KL}(P_{ij} ||Q) = \sum_{m=1}^{k} p_{ij}(\phi_m) \delta_{ij} \log \frac{p_{ij}(\phi_m) \delta_{ij}}{q(\phi_m)h}
\label{eq:pq_bins2}
\end{align}
The summation would include several zero terms of the type $0 \log \frac{0}{q(\phi_m)h}$, corresponding to the empty bins of the histogram for $P_{ij}$. The only non-zero terms are therefore the ones in~\eqref{eq:KL_div_loc2}.

\begin{figure*}[ht!]
	\centering
	\subcaptionbox{PMF of the common baseline Q\label{fig_comparison-sizes:baseline}}{\centering
		\includegraphics[width=0.32\textwidth, angle = 0]{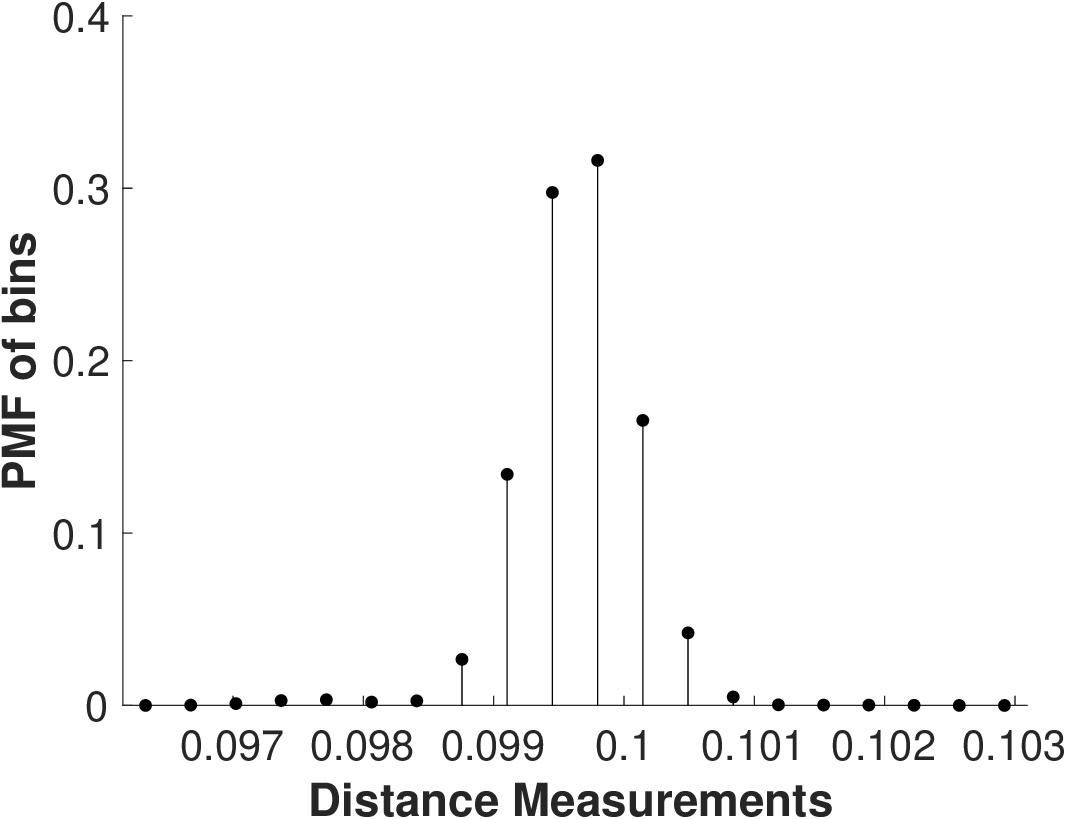}}
   \hfil
	\subcaptionbox{PMF of smaller anomaly data\label{fig_comparison-sizes:small}}{\centering
		\includegraphics[width=0.32\textwidth, angle = 0]{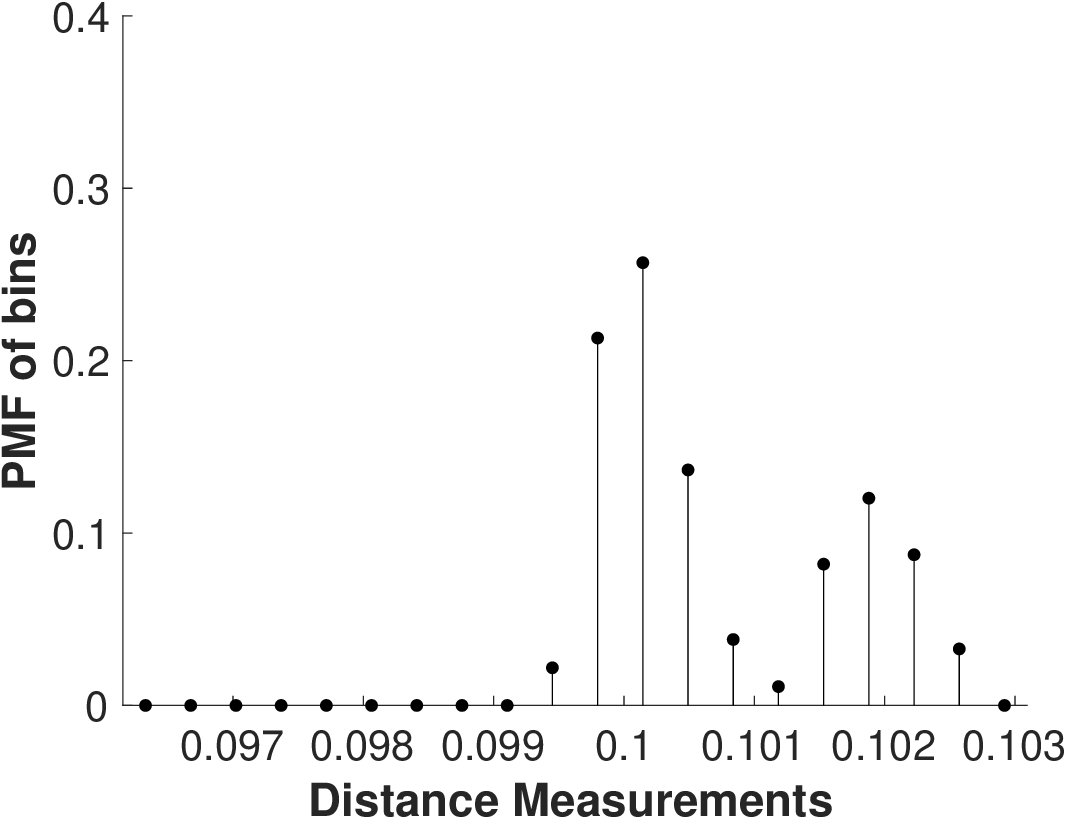}}
	\hfil
	\subcaptionbox{PMF of larger anomaly data\label{fig_comparison-sizes:large}}{\centering
		\includegraphics[width=0.32\textwidth, angle = 0]{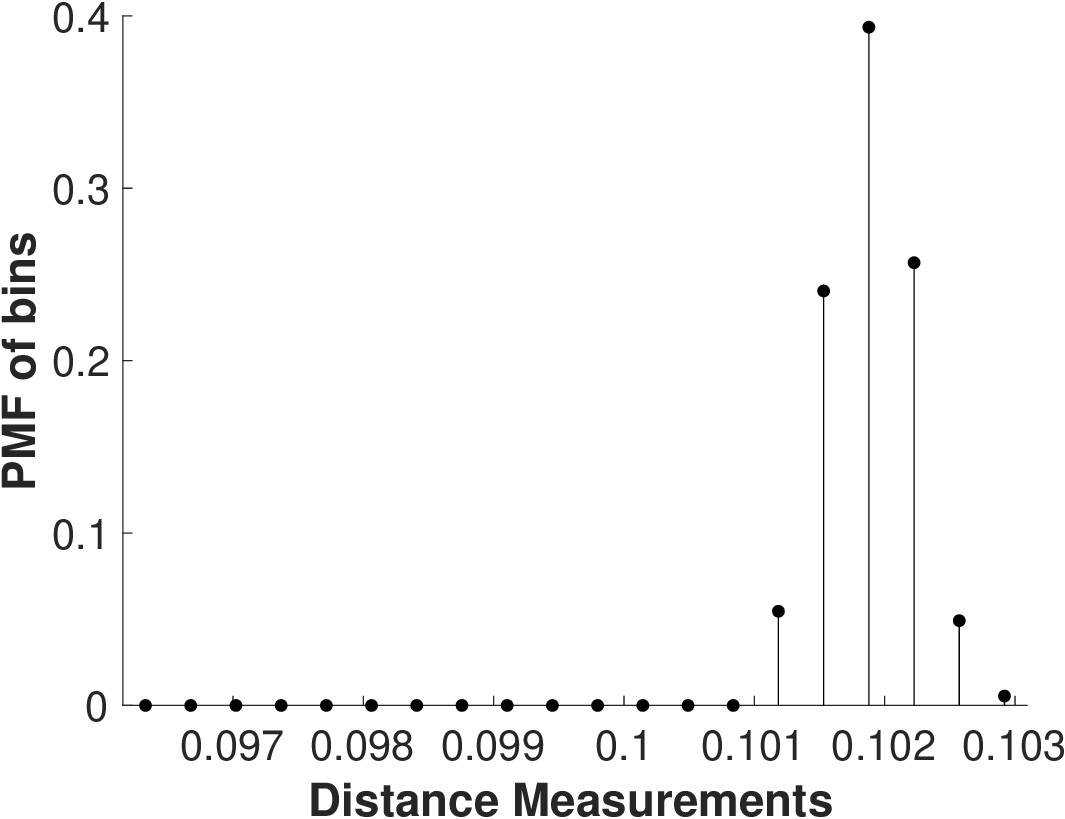}}
              \caption{Comparison between the probability mass functions of two anomalies with different sizes and baseline distribution~$Q$.}
	\label{fig_comparison-sizes}
\end{figure*}

By repeating the process, each data point $\phi_{ij}$ can be mapped to a local KLD value. The pseudocode of this algortithm is provided in Algorithm~\ref{algortihm_kld}. Given the local nature of the KLD map as conferred by the window size around each original data point, the algorithm is able to spatially differentiate different data points. The three probability distributions in Fig.~\ref{fig_comparison-sizes} refer to two points on different data sets from the same proximity sensor and their common baseline, in which anomalies of different sizes were present. The PMF of the data point on the larger anomaly is more narrowly distributed around the sample point representing the anomaly value, which results in a larger discrepancy from the baseline distribution $Q$ in Fig.~\ref{fig_comparison-sizes:baseline}. As we will numerically verify in Section~\ref{results}, anomalies of larger sizes correspond to larger KLD values, establishing therefore an immediate intuitive interpretation of the KLD filtered data.

%

\begin{algorithm}
	
	\SetAlgoLined
	\KwIn{horizontal position max $M$; vertical position max $N$; numer of bins $k$ \& $K$; window size $w$ \& $l$; raw sensory dataset $\mathbb{S}$}
	\KwOut{$KLD(i,j)$ }
	
	
	Calculate PMF for $Q$ by building a $k$-bin histogram from $\mathbb{S}$ \;
	
	\For{$i=1;i \le M; i++$}{
		
		\For{$j=1;j \le N; j++$}{
			
			calculate PMF for $P_{ij}$ by building a $K$-bin histogram from $\Phi_{ij} $ \;
			
			\For{$n=1;n \le K; n++$}{
				
				\eIf{$P(\Phi_{ij}(n))=0$ }{
					
					$kld_ n= 0$\;
					
				}{
					
					$kld_n =  p_{ij}(\phi_n) \delta_{ij} \log \frac{p_{ij}(\phi_n) \delta_{ij}}{q(\phi_n)h} $ eq.~\eqref{eq:pq_bins2}\;
				
				}
				
				$KLD(i,j) = KLD(i,j) + kld_n$\;
				
			}
		}
	}
	
	\caption{Local KLD calculation for each data point}
	
	\label{algortihm_kld}
	
\end{algorithm}

\section{Results and Discussion} \label{results}

\subsection{Setup: A Pipeline Non-Destructive Inspection Case Study}

The proposed KLD filter can be applied for the online detection of anomalies and the estimation of their relative sizes in a large variety of applications and with many types of sensors. In this work, we illustrate its performance in a pipe non-destructive inspection-like environment. To that end, a laser proximity sensor is used to scan the inner surface of a \SI{2}{\milli\meter}-thick pipe with an internal diameter of \SI{400}{\milli\meter}, such that one sample is taken each $(\SI{1}{\milli\meter}, \ang{1})$ of the inner pipe's cylindrical coordinates. The sensor is oriented radially from the pipe's inner surface within a nominal proximity of \SI{100}{\milli\meter}. It has a proximity sensing resolution of \SI{0.1}{\milli\meter} and is characterized by a signal-to-noise ratio (SNR) of \SI{50}{\decibel}, which is typical in commercial sensors, as studied in~\cite{designs2015proximity}.

\begin{figure}[!htb]
\centering
\includegraphics[scale=0.9]{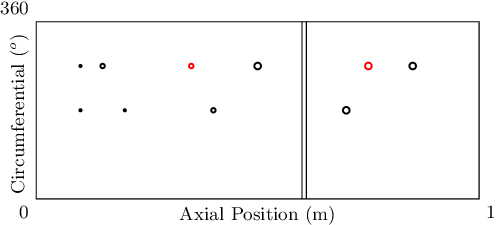}
\caption{Schematic of the anomalies on the pipeline, including ten holes represented by the circles and a welding strip represented by two vertical lines.}
\label{fig_pipe}
\end{figure}

A total of 10 anomalies are distributed over the pipe surface in the shape of circular through holes and blind holes of different diameters. Through holes (or simply holes) cut all through the pipe surface, whereas blind holes penetrate only \SI{1}{\milli\meter} through the surface (that is half of the pipe's thickness). Both types of holes can come in different diameters of 5, 10, and \SI{15}{\milli\meter}. The anomalies are schematized in a two-dimensional representation of the pipe in Fig.~\ref{fig_pipe}, where holes and blind holes are marked by black and red circles, respectively. It is worth mentioning that this test setup is compliant with the standards followed in the gas pipeline non-destructive inspection industry~\cite{spinello_entropy_2011}.

In the case of metal pipes, such as the ones used to transport gas for example, welding strips often appear at the junctions of portions of long pipelines. Therefore, it is important to properly identify them and not mistaken them as anomalies. To include this case in our testing, a welding strip of \SI{10}{\milli\meter} width and \SI{2}{\milli\meter} depth is added to the test pipe. It is marked by two vertical stripes in Fig.~\ref{fig_pipe}. 

\subsection{Results}\label{subsec:results}


%

A study is conducted to decide on the KLD parameter values. Scanning the pipe led to 67 different values of raw sensor readings. As such, the number of bins $k$ of $Q$ is set to 67 to assign every outcome a probability to reveal the distribution of the entire data set. For the distribution $P_{ij}$, the number of bins $K$ is set to 60 given that the window size parameters are taken as $w=1$ and $l=60$. 
It was observed in simulation that by setting the number of bins considerably lower than this results in a significant degradation of performance.

\begin{figure}[h!]
  \centering
  \includegraphics[width=0.99\columnwidth, angle=0]{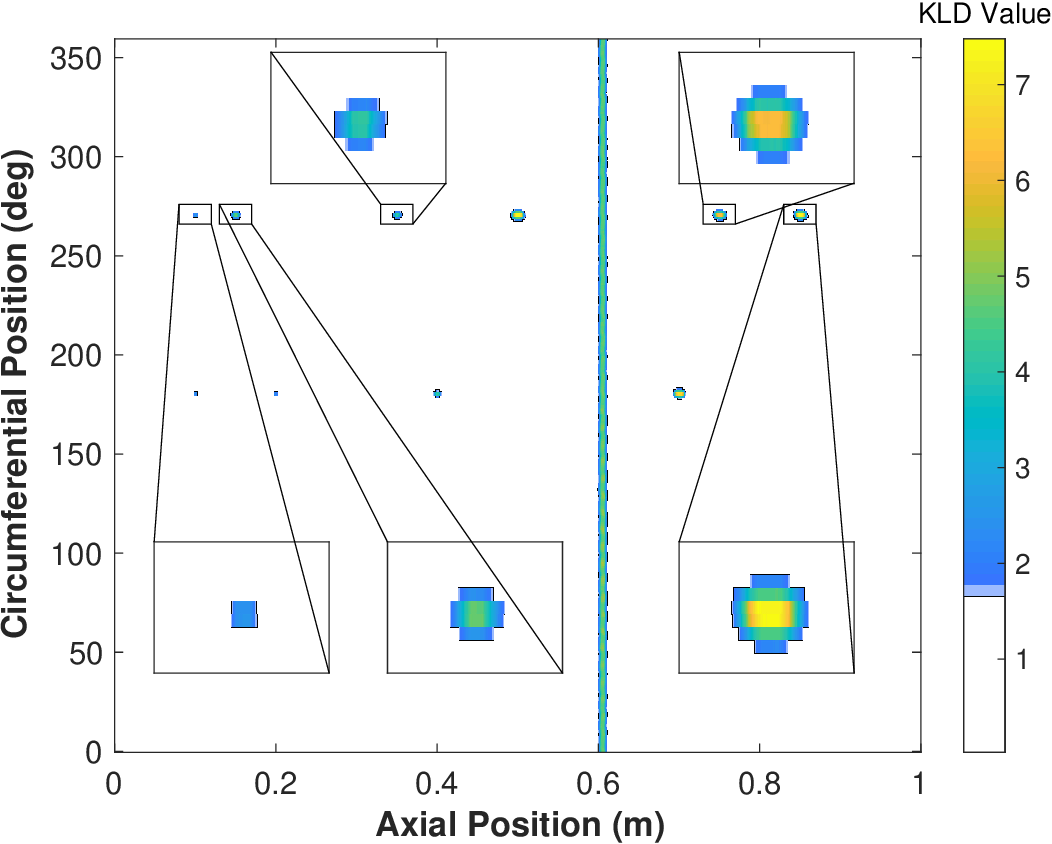}
  \caption{KLD filter output. Anomalies are magnified using the same scale to preserve their relative size.}
  \label{fig_modelfree_result}
\end{figure}

The output of the proposed KLD filter is presented in the form of a color map, as shown in Fig.~\ref{fig_modelfree_result}. All the defects and their locations are successfully detected with no false positives or false negatives, with the background noise effectively filtered out. The true location of the anomalies is unknown to the algorithm, and it is used only to assess the accuracy at detection. The relative sizes of the anomalies are properly marked through their axial diameters on the color map. The colors in each anomaly, which reflect their KLD values, indicate the relative depth of the defect. The figure clearly shows the welding strip as well, which cannot be mistaken for an anomaly. 

Although the current study is more geared towards computing the relative sizes of the anamolaies (with respect to each other) rather than calculating their true sizes, the latter may be possible with a proper calibration. However, more research is still needed along that line to achieve conclusive results.

\subsection{Filter's Window Size Influence: A Qualitative Study}\label{subsec:parameters}

\begin{figure}[htp]
  \centering
  \subcaptionbox{\label{fig_para_window:small-window}}{\centering
    \includegraphics[width=0.99\columnwidth, angle = 0]{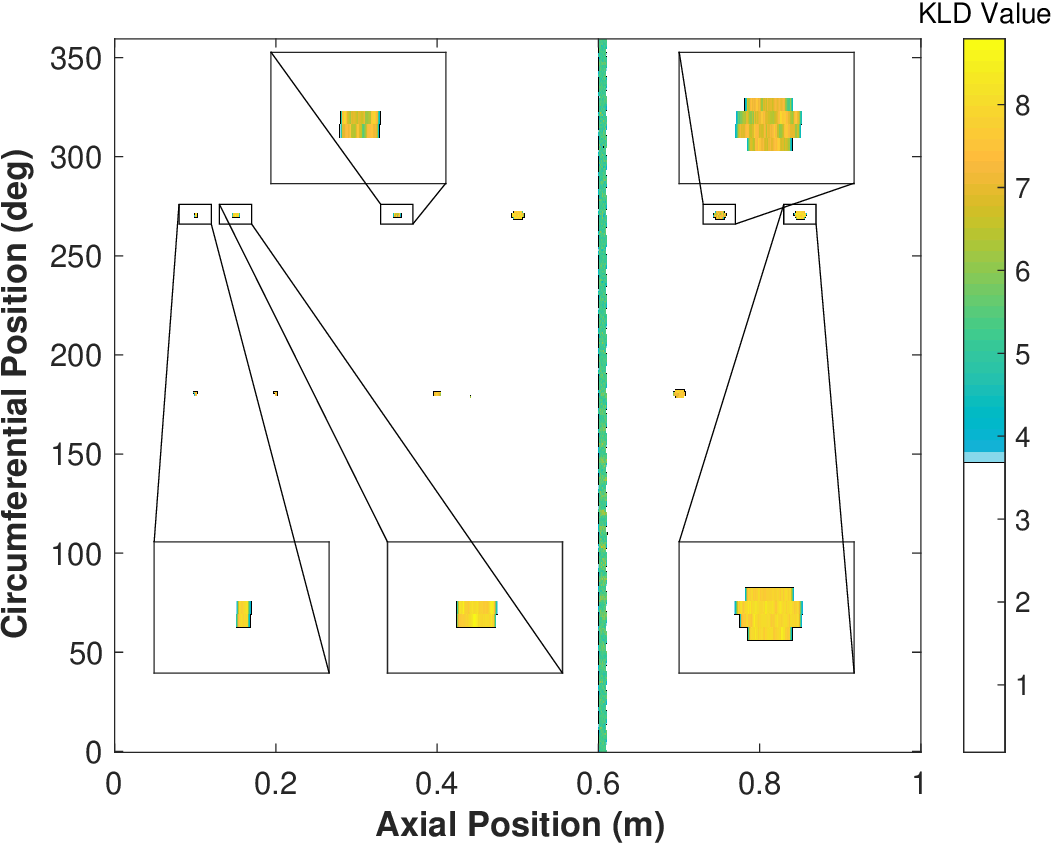}}%
  \\[2ex]
  \subcaptionbox{\label{fig_para_window:large-window}}{\centering
    \includegraphics[width=0.99\columnwidth, angle = 0]{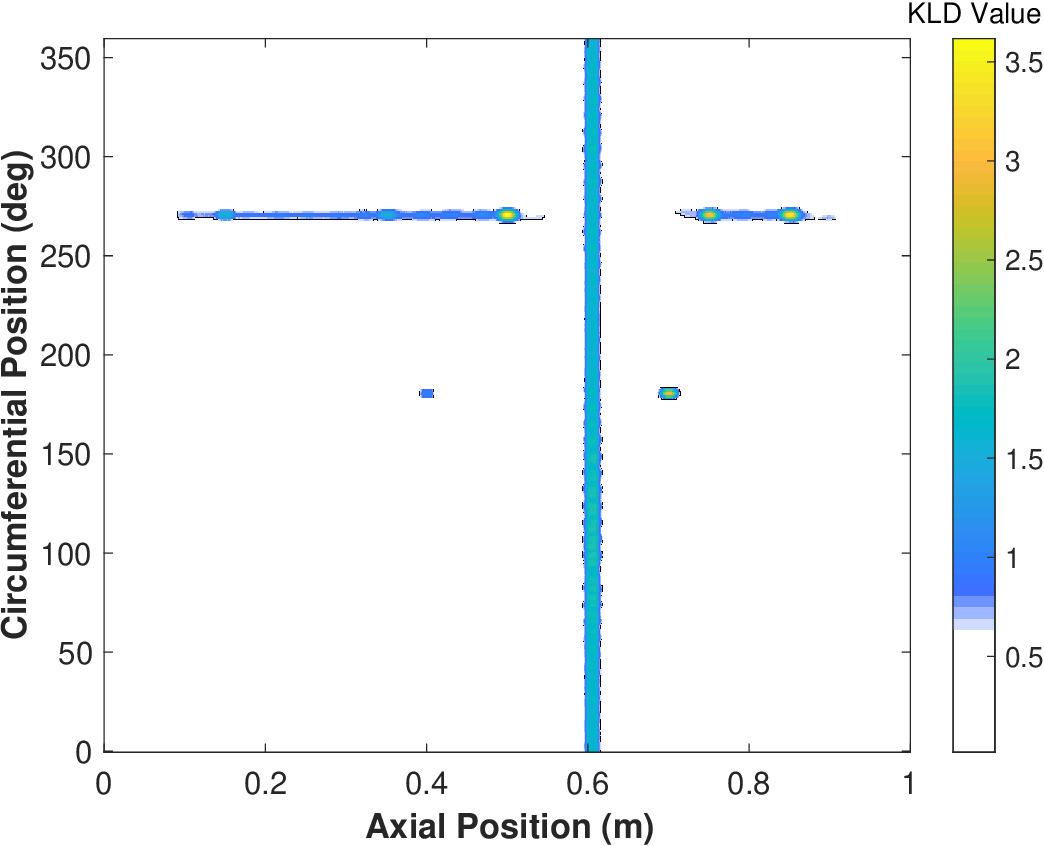}}%
  \caption{KLD filter output with inappropriate window sizes.
    \subref{fig_para_window:small-window}~with $(l=10,w=0)$, the filter cannot relay proper relative anomaly depths.
    \subref{fig_para_window:large-window}~with $(l=200,w=3)$, some healthy regions around the defects are misidentified as anomalies while some small anomalies are not detected at all due to the excessively large window size.}
  \label{fig_para_window}
\end{figure}

The window size is crucial to confer locality to the filter, and ultimately being able to spatially distinguish features embedded in the data set. The window needs to be sufficiently large to include enough data points around a given sample to smooth out eventual local fluctuations. Yet, it has to be sufficiently small to represent relevant trends. For the class of examples considered here, the window size is also important to determine the relative size, especially the depth, of the anomaly, which can be affected by the filter's spatial resolution. Choosing a too small window size makes local posteriors $P_{ij}$ very similar regardless of the stretch of the anomalies, preventing the algorithm from properly assessing their relative sizes. This is clearly shown in Fig.~\ref{fig_para_window}. A narrow window leads to no cross-channel correlation, which results in an inconsistent computation of the anomaly relative depth. In Fig.~\ref{fig_para_window:small-window}, for instance, blind and through holes are deemed to have comparable depths despite the significant difference between them. On the other, a too large window results in identifying some of the healthy regions contiguous to the holes as anomalies, due to spurious spatial correlations, as revealed in Fig.~\ref{fig_para_window:large-window}. 

The effect of the filter locality parameters $w$ and $l$ illustrated in Fig.~\ref{fig_para_window} can be used as design guideline to set the resolution of the filter to detect classes of anomalies with characteristic and/or desired spatial sizes. For data sets with unknown anomalies, the algorithm could be first calibrated on data sets with known anomalies that are of detection interest, by setting the window size based on the area of the anomalies of interest.

\subsection{Assessment of the approximation in the baseline distribution}

To assess the validity of the approximation of using the entire data set that includes the anomalies to build the baseline distribution, we consider the discrepancy introduced by the adoption of this distribution versus adopting a baseline distribution from purely noisy data. As discussed in the previous section, our choice has the advantage of avoiding indeterminate terms in the calculation of the KLD, but it introduces an approximation that is expected to be valid when the anomaly data is relatively small as compared to the size of the entire data set. Let $Q_A$ and $Q_N$ respectively be the baseline distribution from the all data set (including anomalies) and from purely noisy sensing data, for example obtained from raw data series of the types in Fig.~\ref{fig:rawData}. For a given data point, \eqref{eq:KL_div_loc} gives the map to the local KLD which can be manipulated to 
{\allowdisplaybreaks \begin{align}\label{eq:KL_div_err}
D_{\rm KL}(P_{ij} ||Q_A) &= \sum_{\phi \in \Phi_{ij}} P_{ij}(\phi) \log \frac{P_{ij}(\phi)}{Q_A(\phi)} \nonumber
\\
&= \sum_{\phi \in \Phi_{ij}} P_{ij}(\phi) \log \frac{P_{ij}(\phi)}{Q_N(\phi)} \frac{Q_N(\phi)}{Q_A(\phi)} \nonumber
\\
&= D_{\rm KL}(P_{ij} ||Q_N) +  \sum_{\phi \in \Phi_{ij}} P_{ij}(\phi) \log \frac{Q_N(\phi)}{Q_A(\phi)}
\end{align}}%
This allows to define the error $\delta_A = D_{\rm KL}(P_{ij} ||Q_A) - D_{\rm KL}(P_{ij} ||Q_N)$ induced by the adoption of the baseline $Q_A$ instead of the baseline $Q_N$. In view of \eqref{eq:KL_div_err} the error can be computed by
\begin{align}
\delta_A = \sum_{\phi \in \Phi_{ij}} P_{ij}(\phi) \log \frac{Q_N(\phi)}{Q_A(\phi)}
\end{align}
Note that when the window size around the data point includes the entire data set we have $\delta_A = - D_{\rm KL}(Q_A ||Q_N)$, which is a measure of discrepancy between the two baseline distributions. Since we want to check the influence of the approximation $Q_N \simeq Q_A$ on the KLD filter, we plot the sensitivity parameter $\delta_A/D_{\rm KL}(P_{ij}||Q_A) = 1 - D_{\rm KL}(P_{ij}||Q_N)/D_{\rm KL}(P_{ij}||Q_A)$ by generating data sets with increasing percentage of anomaly data (increasing number of anomalies of the same type considered in Section~\ref{subsec:results}), while keeping the distribution $P_{ij}$ fixed. As discussed at the end of Section~\ref{KL divergence}, the adoption of a baseline distribution with noise only, as it is the case of $Q_N$, requires some ad hoc assumption in the calculation of the KL divergence with discrete distributions since anomaly data points have zero probability in $Q_N$. Following \cite{cover_elements_2006}, we adopt the convention $P_{ij} \log \frac{0}{Q_A} = 0$, discarding the terms in the summation for which the distribution $Q_N$ evaluates to zero. When the anomaly data is relatively rare the distributions $Q_A$ and $Q_N$ are very similar and the sensitivity parameter approaches zero, whereas when the percentage of anomaly data significantly increases the parameter grows significantly since the divergence between $P_{ij}$ and $Q_A$ decreases as they become more similar.

\begin{figure}[h!tp]
	\centering
	\includegraphics[scale=0.36, angle = 0]{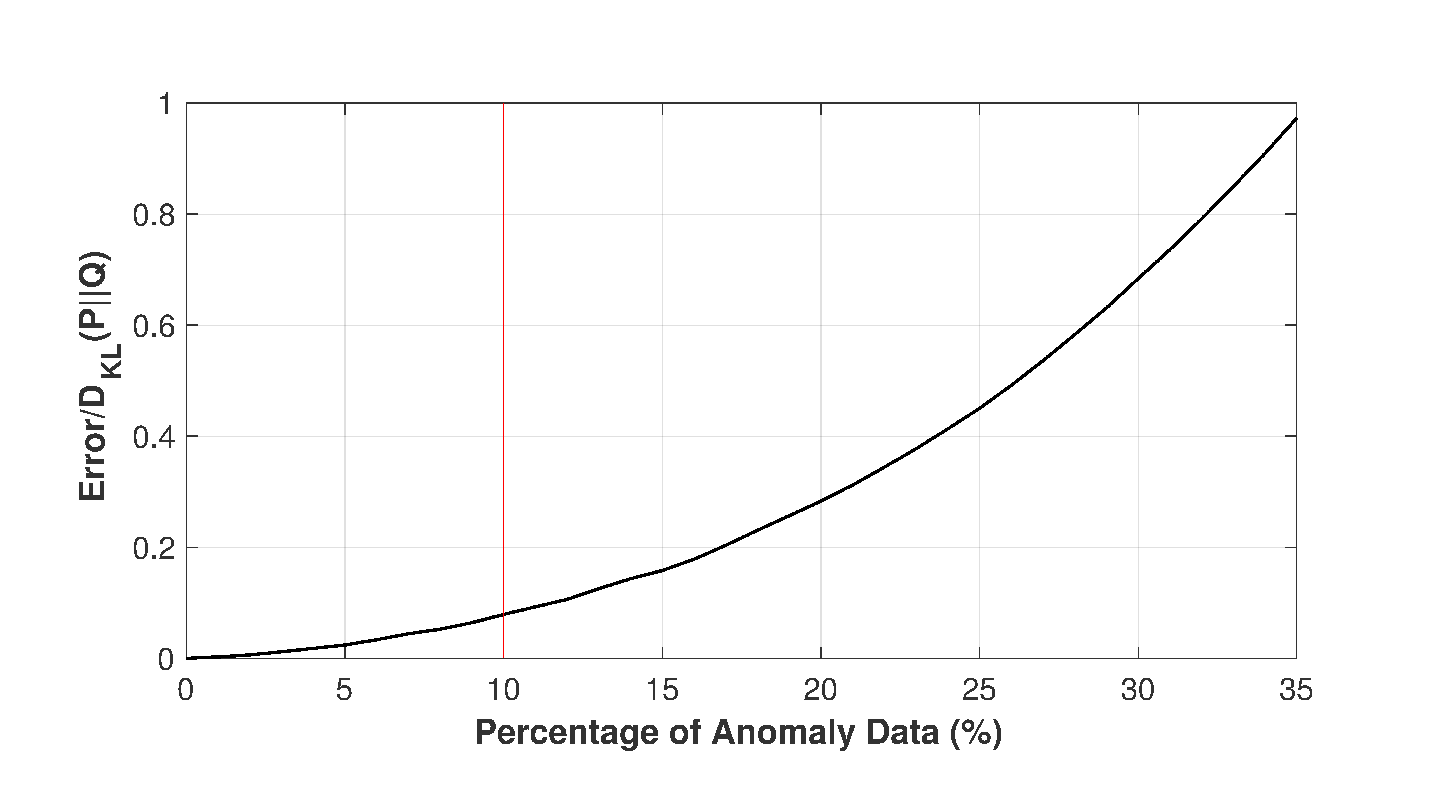}
	\caption{Sensitivity $\delta_A/D_{\mathrm{KL}}(P_{ij}||Q_A)$ versus percentage of anomaly data in the entire data set.}
	\label{fig:KLD approx}
\end{figure}

Results in Fig.~\ref{fig:KLD approx} are calculated for a data point with distribution $P_{ij}$ which includes about 50\% of noisy data (a data point in the neighbour of an anomaly). The vertical line marks the point at which the results in Section~\ref{subsec:results} are calculated (about 10\% of anomaly data in the baseline distribution), showing that the influence is about $5\%$ for a data point of this type. Considering that data series with anomaly are typically dominated by noise, this analysis suggests that the approximation that we adopted to avoid indeterminate addends in the calculation of the KLD filter can be viable for this class of applications. 

For data points at the center of the anomalies, the error $\delta_A$ approaches zero since the baseline $Q_N$ does not include anomaly data (zero probability), and therefore a large amount of terms in the summation in \eqref{eq:KL_div_err} are zero. Therefore, the influence of the approximation on the KLD filter is negligible on the critical regions that include anomalies when the filter is sufficiently local. 

\section{Conclusion} \label{conclusion}

We proposed an anomaly detection algorithm operating on noisy data series that could be typically produced by a large class of sensing devices and applications. The algorithm maps the raw sensory data to a local KLD value that measures the discrepancy between two probability distributions. Leveraging on this feature, we build a baseline probability distribution (as a prior) from the whole data set. A posterior probability distribution is computed form a subset of the data centered around a given sample. The discrepancy between the two, as measured by the corresponding KLD, is a local measure of difference in information content between the data in the neighborhood of the given sample and the overall data set which is dominated by noise around the sensor's expected measurement. The KLD filter's output is obtained by repeating this procedure for every sample point of the raw data.

The algorithm is tested in a setup that is compliant with the standards followed in the gas pipeline inspection industry. It successfully detected all the anomalies, their locations, relative surface areas and depths, without false negatives or false positives. To achieve such a performance, it is first necessary to tune some of the filter's most influential parameters. To this end, a qualitative study on the effect of these parameters on the filter's performance has also been included. 

\begin{acknowledgment}
This work was supported in part by NSERC under Grants 2014-06512 and 2016-05783.
\end{acknowledgment}


%
% ...

\bibliographystyle{asmems4}

\bibliography{KLDbiblio}



\end{document}